\documentclass[manuscript,screen]{acmart}

\AtBeginDocument{%
  }

\setcopyright{acmlicensed}
\copyrightyear{2025}
\acmYear{2025}
\acmDOI{XXXXXXX.XXXXXXX}

\acmConference[Conference acronym 'CHI25]{The ACM (Association of Computing Machinery) CHI conference on Human Factors in Computing Systems}
\acmISBN{978-1-4503-XXXX-X/18/06}

\usepackage{alltt}

\usepackage[toc,page]{appendix}

\begin{document}

\title{Enhancing Mental Health Support through Human-AI Collaboration: Toward Secure and Empathetic AI-Enabled Chatbots}

\settopmatter{printacmref=false}

\author{Rawan AlMakinah}
\email{ralmakinah@albany.edu}
\orcid{0009-0006-4682-3848}
\affiliation{
  \institution{AI in Complex Systems Lab, University at Albany, SUNY}
  \city{Albany}
  \state{New York}
  \country{USA}
}

\author{Andrea Norcini-Pala}
\email{andrea.norcinipala@downstate.edu}
\affiliation{
  \institution{SUNY Downstate Health Sciences University}
  \city{Brooklyn}
  \state{New York}
  \country{USA}
}

\author{Lindsey Disney}
\email{ldisney@albany.edu}
\affiliation{
  \institution{University at Albany, SUNY}
  \city{Albany}
  \state{New York}
  \country{USA}
}

\author{M. Abdullah Canbaz}
\email{mcanbaz@albany.edu}
\affiliation{
  \institution{AI in Complex Systems Lab, University at Albany, SUNY}
  \city{Albany}
  \state{New York}
  \country{USA}
}

\renewcommand{\shortauthors}{Trovato et al.}

\begin{abstract}
Access to mental health support remains limited, particularly in marginalized communities where structural and cultural barriers hinder timely care. This paper explores the potential of AI-enabled chatbots as a scalable solution, focusing on advanced large language models (LLMs)—GPT v4, Mistral Large, and LLama V3.1—and assessing their ability to deliver empathetic, meaningful responses in mental health contexts. While these models show promise in generating structured responses, they fall short in replicating the emotional depth and adaptability of human therapists. Additionally, trustworthiness, bias, and privacy challenges persist due to unreliable datasets and limited collaboration with mental health professionals. To address these limitations, we propose a federated learning framework that ensures data privacy, reduces bias, and integrates continuous validation from clinicians to enhance response quality. This approach aims to develop a secure, evidence-based AI chatbot capable of offering trustworthy, empathetic, and bias-reduced mental health support, advancing AI’s role in digital mental health care.
\end{abstract}



\maketitle 
\section*{Acknowledgement}
This paper has been submitted for review to the Association of Computing Machinery (ACM) CHI25 conference on Human Factors in Computing Systems.

\section{Introduction}
In recent years, the global mental health crisis has reached unprecedented levels, with the World Health Organization (WHO) reporting a 25\% increase in anxiety and depression during the COVID-19 pandemic alone \cite{world_health_organization_covid-19_2022}. Mental health issues have become a significant global concern, with millions of cases going undetected and untreated due to various structural and attitudinal barriers \cite{andrade_barriers_2014, clement_what_2015, fukuda_mental_2016}. Structural barriers, such as limited access to mental health services and high costs, prevent individuals from receiving timely and effective care \cite{daley_preliminary_2020}. Attitudinal barriers, including the stigma surrounding mental health and lack of awareness, further exacerbate the problem, leading many to avoid seeking help even when it is available \cite{andrade_barriers_2014, yazdavar_analyzing_2016}. As a result, untreated mental health conditions often worsen, leading to adverse physical, economic, and emotional outcomes \cite{vammen_emotional_2016, alsagri_machine_2020}.

To address these challenges, digital mental health solutions have emerged as a promising alternative to traditional mental health services. The rapid development of technologies, increased internet accessibility, and widespread use of smartphones have paved the way for innovative approaches to mental health care \cite{daley_preliminary_2020}. Among these solutions, AI-enabled chatbots have gained considerable attention due to their ability to provide cost-effective, scalable, and accessible mental health support. Chatbots such as Woebot \cite{fitzpatrick_delivering_2017}, Wysa \cite{inkster_empathy-driven_2018}, and ViTalk \cite{daley_preliminary_2020} have demonstrated their potential in offering empathic interaction and cognitive-behavioral therapy (CBT), a widely used intervention in treating mental health challanges like anxiety and depression \cite{lee_large_2024, fitzpatrick_delivering_2017}. These AI-enabled systems not only mimic human-like interactions but also offer anonymity and 24/7 availability, making them an appealing option for individuals hesitant to seek traditional mental health services due to stigma or other concerns \cite{balcombe_ai_2023, haque_overview_2023}.

The rise of large language models (LLMs), such as GPT-3 and GPT-4, has further revolutionized the field of AI and its application to mental health care \cite{brown_language_2022}. These models, trained on vast amounts of text data, possess the ability to generate coherent, human-like responses in real-time, making them well-suited for use in chatbots aimed at providing psychological support. LLMs can be fine-tuned to understand and respond to specific mental health needs, including crisis intervention and emotional support, allowing them to play a critical role in mental health first aid \cite{menezes_use_2019, vaidyam_chatbots_2019}. By simulating natural conversations, these AI systems have the potential to fill the gaps in mental health services, particularly in underserved communities where access to mental health professionals is limited \cite{fitzpatrick_delivering_2017}.

However, despite the promising potential of AI-enabled chatbots in mental health, significant challenges remain. Ethical and practical concerns regarding the reliability, security, and trustworthiness of these systems have hindered their full integration into healthcare frameworks \cite{balcombe_ai_2023, casu_ai_2024}. Ensuring that chatbots are not only effective but also empathetic—able to recognize and respond to users’ emotional states with reduced bias and enhanced security is critical for gaining public trust and promoting widespread adoption. Moreover, the collaboration between data scientists and mental health practitioners is essential to address these concerns and ensure that AI-enabled chatbots are aligned with evidence-based mental health practices \cite{graham_artificial_2019}.

In this paper, we focus on analyzing large language models (LLMs) for their capacity to comprehend and respond to emergency mental health scenarios, specifically focusing on mental health first aid. Following this analysis, we propose a conceptual design for an AI-enabled chatbot that integrates state-of-the-art advancements in AI, human-computer interaction (HCI), and mental health care practices. This interdisciplinary approach is critical to address the growing mental health needs, where AI and HCI can help overcome accessibility and stigma-related barriers. By adopting a human-AI collaboration approach, we aim to build a trustworthy, secure, and reduced bias AI-enabled chatbot, with mental health professionals in the loop to ensure ethical and practical challenges are met. Our framework also suggests the integration of federated learning and encryption techniques to further enhance the security and reliability of AI chatbots in mental health care systems. 

\section{Methodology}

\subsection{Knowledge/Comprehension Measurement}

Prompt engineering is the process of carefully crafting the input provided to an AI model in order to generate desired outputs \cite{velasquez-henao_prompt_2023}. In the context of using LLMs like ChatGPT, prompt engineering plays a crucial role in guiding the model to respond appropriately and effectively, especially in specialized or sensitive domains like mental health. By tailoring prompts to include specific instructions, context, or even role-based personas, users can control the behavior of the model and improve its performance on complex tasks \cite{schreiber_pattern_nodate}. This is particularly important when utilizing off-the-shelf AI tools, as these models do not inherently understand the full context of the problem they are addressing without clear guidance.

\begin{figure}[h]
  \centering
  \includegraphics[width=\linewidth]{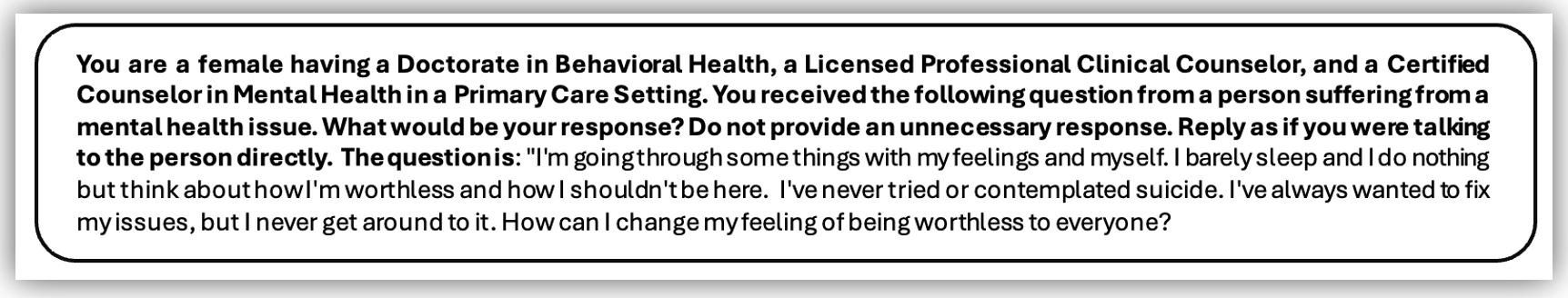}
  \caption{LLMs Prompt Example}
  \Description{LLMs Prompt Example}
  \label{prompt}
\end{figure}

For example, when generating therapeutic responses, instructing the model to act as a professional counselor through prompt engineering ensures that the tone, language, and structure of the response are aligned with professional standards. Without such prompts, the model might produce generic or inappropriate responses, potentially diminishing the effectiveness of AI-assisted support - or worse, causing harm. Figure~\ref{prompt} demonstrates an example of such a prompt, where specific instructions were given to simulate responses as if from a licensed mental health professional.

To evaluate the capabilities of AI-enabled chatbots in replicating the expertise of mental health professionals, we adopted a comparative approach using two large language models (LLMs): ChatGPT and Mistral. The goal was to measure the models' ability to comprehend and provide supportive responses to complex mental health issues.

We developed specific \textit{prompts} to include the \textit{persona (or profiles)} for the LLMs to simulate the behavior of mental health professionals. These profiles included contextual information designed to reflect the therapeutic language and professional demeanor of licensed counselors. The purpose of creating such profiles was to ensure that the LLMs responded with the level of professionalism and empathy required in mental health interactions. Figure \ref{prompt} presents an example of a prompt crafted for one of the data points, incorporating the therapist’s background, education, and the patient’s condition. It is important to note that each prompt used in our analysis featured a unique persona tailored to the therapist’s qualifications and experience, as noted in the dataset. This personalized approach aimed to enhance the LLMs’ ability to communicate empathetically and utilize appropriate mental health terminology, drawing from the ‘CounselChat’ dataset \cite{bertagnolli_counsel_2023}.

\subsection{Question Set Selection}

The CounselChat dataset consisted of user-submitted questions on a wide range of mental health topics, with responses provided by certified counselors. For this study, we focused on questions related to eight crisis-related categories: anxiety, depression, grief and loss, self-harm, stress, anger management, trauma, and military issues. This filtering process resulted in approximately 700 questions.

To refine the analysis further, we identified the ten most-viewed questions on the platform, which predominantly fell within the categories of anxiety, depression, anger management, and military issues. These top 10 questions were then selected for manual analysis.

\subsection{Analysis of Response Generation and Comparison for Top 10 Questions}

For this part of the analysis, we utilized two of the most popular models: OpenAI’s \textit{ChatGPT v4} and its open-source competitor, \textit{Mistral Large}. The selected questions were input into both models, using the respective prompts shown in Figure \ref{prompt}. The responses generated by the LLMs were compared to those of human counselors, with a focus on assessing strengths and weaknesses in empathy, comprehension, and therapeutic value.

\subsection{Preliminary Qualitative Analysis of Therapist vs LLM Generated Responses}

We began our analysis with a manual comparison of responses generated by ChatGPT and Mistral against those provided by human therapists. This preliminary qualitative assessment allowed us to explore the differences in tone, empathy, and overall communication style. It quickly became evident that while AI-generated responses can be coherent and professional, they often lack the nuanced emotional intelligence found in human responses.

Key distinctions emerged from this analysis: (1) The AI responses tended to have a polished but somewhat generic tone, delivering psychoeducational content in a way that felt mechanical and impersonal. This lack of a “human touch” made them easily identifiable as machine-generated. (2) In contrast, therapist responses were more concise and dialogic, incorporating questions aimed at fostering engagement and building a ‘working therapeutic relationship.’ This approach aligns with established mental health practices that emphasize checking in with the client to create a more interactive and supportive dialogue, as opposed to the didactic and instructional style often seen in AI responses. (3) Despite these differences, certain similarities were present. All responses—whether from the therapist, ChatGPT, or Mistral—featured empathetic statements and included essential psychoeducational content related to anxiety, as demonstrated in Appendix \ref{Appendix_Output}.

These qualitative findings prompted the need for a more detailed quantitative analysis to further explore the extent of these differences, which we carried out using cosine similarity.

\subsection{Quantitative Analysis of Therapist vs LLM Generated Responses}

To quantify the differences between the LLM-generated responses and those of human therapists, we conducted a Cosine Similarity Analysis. This technique compares two sets of text data—therapist vs. LLM responses—by converting them into vector form and measuring the cosine of the angle between them. A cosine similarity score of 1 indicates identical vectors, while a score of 0 represents completely dissimilar vectors \cite{bhattacharjee_sentiment_2015}.

For this analysis, we employed Term Frequency-Inverse Document Frequency (TF-IDF) to calculate cosine similarity. TF-IDF is particularly well-suited for this task because it assigns weight to words based on their importance in individual documents (therapist or LLM responses) and their rarity across the entire dataset. In the context of mental health responses, where certain key terms like “empathy” or “anxiety” carry significant meaning, TF-IDF helps capture both the specific therapeutic language used by human therapists and the critical elements of AI-generated responses. This ensures that the comparison isn’t simply based on word counts, but rather on the significance of the terms used in context. The formula for cosine similarity is:

\[
\text{Cosine Similarity} = \frac{\mathbf{A} \cdot \mathbf{B}}{\|\mathbf{A}\| \|\mathbf{B}\|}
\]

\begin{figure}[h]
  \centering
  \includegraphics[width=\linewidth]{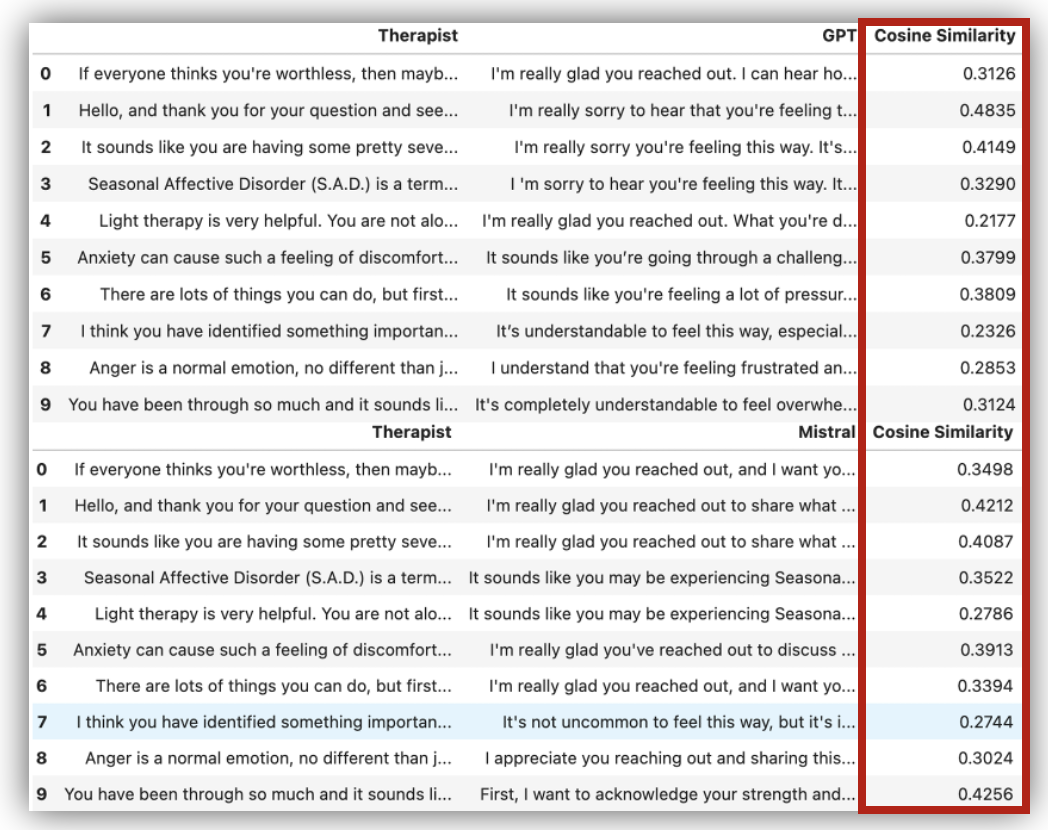}
  \caption{Cosine Similarity Analysis of Therapists and LLMs Responses for 10 Questions}
  \Description{Cosine Similarity Analysis of Therapist and LLMs Responses for 10 Questions}
  \label{CSA}
\end{figure}

Using Python, we calculated cosine similarity scores between the therapist responses and those generated by ChatGPT and Mistral for selected questions. The results, shown in Figure ~\ref{CSA}, reveal a clear gap between the LLMs and human therapists. The majority of the cosine similarity scores fall between 0.21 and 0.48, indicating that while the LLMs can generate syntactically correct and coherent responses, they often fail to capture the deeper emotional and empathetic elements present in human therapeutic communication.

Upon closer inspection, ChatGPT’s scores, which range from 0.2177 to 0.4835, reflect a tendency to provide generalized reassurance and support, often lacking the personalization and interactive elements typical of a therapist’s response. Similarly, Mistral’s scores, ranging from 0.2744 to 0.4265, suggest that its responses are more formulaic. While Mistral’s output often adheres to the structure and tone of a professional response, it lacks the flexibility and emotional responsiveness that are critical in a therapeutic setting.

For instance, ChatGPT’s highest similarity score (0.4835) was achieved in a response that echoed supportive language but fell short in delivering the interactive, dialogue-driven communication seen in human therapists. Mistral’s highest score (0.4265) reflected its ability to replicate the format of a therapeutic response, but again, the model struggled with adapting its responses dynamically to the emotional needs of the user.

In conclusion, the lower cosine similarity scores for both LLMs emphasize their limitations in matching the depth of empathy, emotional engagement, and intuition characteristic of human counselors. These findings, in conjunction with the manual qualitative analysis, highlight the ongoing challenge of bridging the gap between AI-generated responses and the nuanced therapeutic expertise that defines human mental health care.

\begin{figure*}[h]
  \centering
  \includegraphics[width=\linewidth]{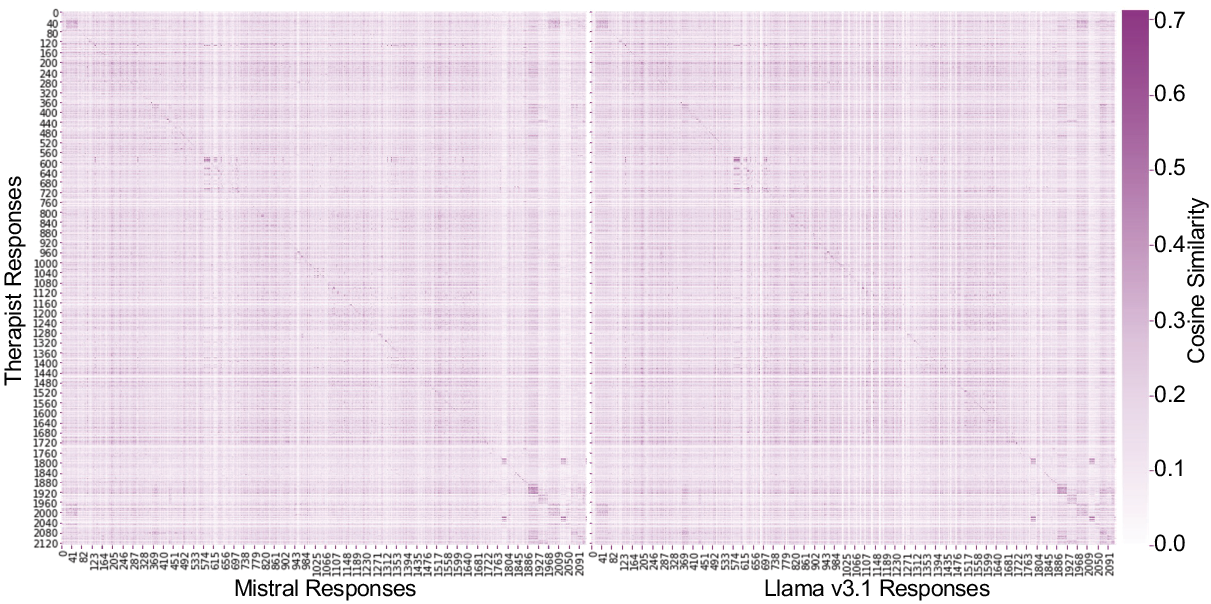}
  \caption{A Heatmap of Cosine Similarity Analysis of Therapist Compared to LLMs Responses(Mistral on Left, LLamav31 on Right)}
  \Description{A Heatmap of Cosine Similarity Analysis of Therapist Compared to LLMs Responses(Mistral on Left, LLamav31 on Right)}
  \label{TherapistvsModels}
\end{figure*}

Next, we expanded our analysis to encompass a larger dataset of 2,129 questions. These questions, all answered by human therapists, were compared to responses generated by two of the most advanced open-source models: Mistral Large and Llama v3.1 13B. Using cosine similarity, we evaluated how well these models replicate the tone, empathy, and therapeutic insight of licensed professionals. The heatmap in Figure ~\ref{TherapistvsModels} visually depicts the similarity between the models’ responses and those of the human therapists, with darker colors indicating higher similarity. This analysis provides a broader view of the models’ performance across a large dataset of therapeutic scenarios.

From the heatmap, it is evident that both Mistral and Llama v3.1 display overall low cosine similarity to the human therapist responses across the dataset of 2,129 questions. The predominance of lighter shades across the map indicates that these models struggle to replicate the nuanced therapeutic communication provided by licensed professionals. Despite their technological sophistication, both models fall short of consistently capturing the depth of empathy, emotional attunement, and interactive conversational style that are critical in mental health support. This aligns with earlier findings where LLMs generated responses that were more structured and syntactically correct but lacked the human touch and emotional complexity.
\begin{figure}[b]
  \centering
  \includegraphics[width=0.7\linewidth]{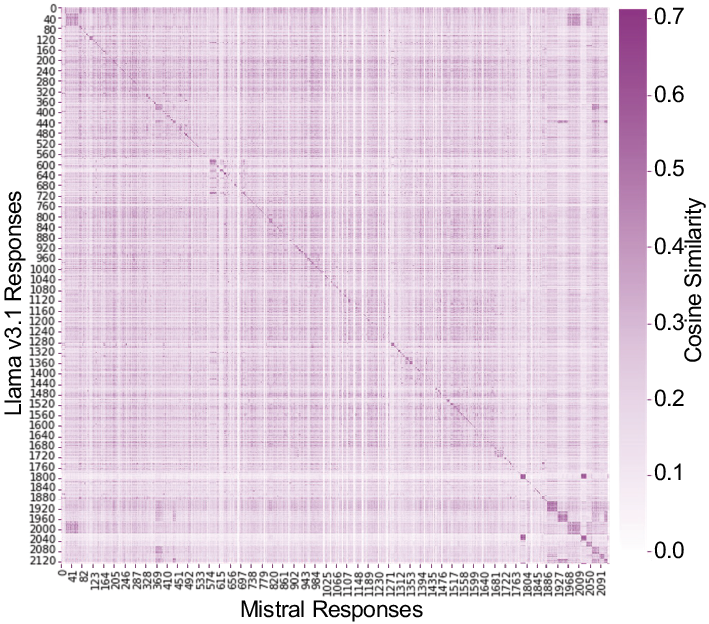}
  \caption{Cosine Similarity Cross Comparison of LLama v3.1 vs Mistral Responses}
  \Description{Cosine Similarity Cross Comparison of LLama v3.1 vs Mistral Responses}
  \label{ModelvsModel}
\end{figure}
There are, however, a few instances where both models show higher cosine similarity, represented by the darker areas in the heatmap. These pockets suggest that in certain cases, Mistral and Llama v3.1 managed to produce responses that were more aligned with the human therapists’ answers, possibly when delivering straightforward psychoeducational content. These higher similarity scores may indicate that the models can effectively mimic the surface-level structure of therapeutic language, particularly in responses that involve more factual or direct information. Nevertheless, these instances are rare, indicating the models’ limitations in consistently generating responses that match the emotional depth and personalization characteristic of human therapists.

Moreover, the heatmap shows no significant difference between Mistral and Llama v3.1 in terms of their performance. Both models exhibit similar patterns of low similarity, suggesting that while they are among the most advanced open-source models, they share common challenges when applied to emotionally charged and complex therapeutic scenarios. This reinforces the idea that while these models are capable of producing relevant content, they often lack the adaptability and emotional intelligence that is central to effective therapy, highlighting the continued gap between AI-generated and human therapist responses.

The heatmap in Figure ~\ref{ModelvsModel} presents a cosine similarity cross-comparison between responses generated by Mistral and Llama v3.1 across the 2,129 questions. The diagonal line of darker shading indicates instances where the same question was input into both models, with relatively higher cosine similarity between their responses. However, for the majority of the responses, the lighter shades across the heatmap reflect generally low similarity, suggesting that while both models are highly advanced, they generate distinct responses for the same questions. This indicates that although Mistral and Llama v3.1 might follow similar linguistic structures, their approach to providing answers varies, resulting in different outputs. The overall variation between these two models underscores their differing methods in handling the same mental health-related questions, despite being advanced in their design and capabilities.

\subsection{Insights and Implications for Human-AI Interaction in Mental Health}

The comparative analysis of LLM-generated responses, both through cosine similarity and qualitative methods, offers valuable insights into the current limitations and potential of AI in interacting with human emotions in mental health contexts. The consistently low cosine similarity scores across models highlight a key shortcoming: while AI models such as Mistral and Llama v3.1 can generate structured, coherent responses, they often lack the deeper emotional attunement and psychological nuance characteristic of human therapists. This gap is especially critical in therapeutic settings, where understanding and addressing complex emotions requires more than delivering factual information—it requires empathy, personalization, and adaptive interaction.

The use of detailed personas and profiles was intended to guide LLMs toward more human-like interactions, helping them adopt the tone and therapeutic language expected of mental health professionals. While these personas aided in shaping the responses toward a more professional demeanor, the cosine similarity analysis reveals that the models still struggled to capture the depth and emotional responsiveness that are critical in therapeutic dialogues. The qualitative analysis further supports this, showing that AI responses, although well-structured, often lacked the emotional sensitivity and adaptive feedback seen in human therapists’ responses. This suggests that current LLMs are more successful at mimicking surface-level professional behavior than at engaging meaningfully with the underlying emotional complexities of clients in distress.

This limitation, however, does not entirely undermine the role of AI in mental health support. While LLMs may not yet be capable of replicating the full range of human emotional intelligence, their ability to provide accurate, structured responses could serve as a foundation for more specialized therapeutic tools. Importantly, the findings underscore the need for continuous human oversight in AI-enabled mental health applications. Without clinical guidance, there is a significant risk that LLMs may deliver responses that, while ‘correct’ in form, fail to meet the emotional needs of users or could even cause harm. Future research must explore whether AI systems can be refined to better detect and respond to emotional cues, or if their role should be limited to more supportive, adjunct functions in a therapeutic setting.

Looking ahead, the implications for Human-AI interaction in mental health are clear. To develop more effective AI systems, future research should focus on improving models’ ability to process not only the linguistic content of user input but also the emotional undertones that are essential in therapeutic exchanges. This may involve integrating more sophisticated emotional recognition capabilities and utilizing multi-modal data, such as voice tone or facial expressions, to provide a fuller understanding of the user’s emotional state. Such advancements would move AI closer to providing responses that align not just with the content but also with the emotional needs of individuals, making AI a more effective tool in mental health care.

\subsection{Feasibility Analysis of Needs for AI-Enabled First Aid Mental Support}

Building on the insights regarding the limitations and potential of current AI models in mental health interactions, we now turn to the feasibility of developing an AI-enabled chatbot that can effectively provide first aid mental health support. To ensure such a system can meet the unique demands of mental health care, we must consider key factors such as trustworthiness, security, bias mitigation, empathy, and privacy.

\begin{figure}
  \centering
  \includegraphics[width=4in]{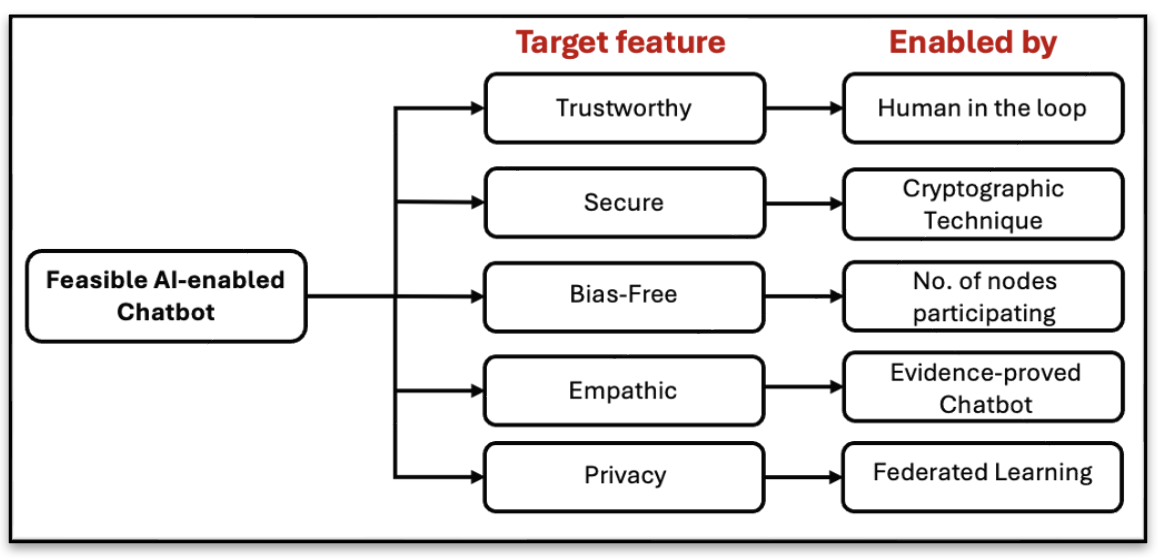}
  \caption{A summary map of the proposed framework features}
  \Description{A summary map of the proposed framework features}
  \label{FeaturseMap}
\end{figure}

The development of an AI-enabled chatbot for mental health support must adhere to a rigorous set of criteria to ensure its effectiveness, reliability, and acceptance by both users and healthcare professionals. As mental health continues to be a growing global concern, exacerbated by the COVID-19 pandemic, the role of digital tools in providing scalable, accessible, and private support has become increasingly important. To design a \textbf{feasible AI-enabled chatbot}, it is essential to address key concerns surrounding trustworthiness, security, bias mitigation, empathy, and privacy, shown in Figure ~\ref{FeaturseMap}.

In the context of mental health, \textbf{trustworthiness} is paramount. Users must be able to trust the chatbot to provide accurate, timely, and ethically sound responses to their needs. This trust can be bolstered by incorporating a \textbf{human-in-the-loop} system, wherein human oversight is used to monitor, validate, and guide the AI’s decisions and interactions. Such oversight ensures that the chatbot behaves in ways that align with human values and ethical standards, mitigating potential harm and fostering user confidence.

Additionally, the chatbot must be \textbf{secure}, especially given the sensitive nature of mental health data. Mental health interactions involve the disclosure of personal, often vulnerable, information. Therefore, robust \textbf{cryptographic techniques} are required to safeguard this data from unauthorized access or misuse. By employing state-of-the-art encryption, the chatbot can ensure that all communications between users and the system remain confidential, protecting users' privacy and enhancing their willingness to engage.

A crucial component of the system’s design is its ability to be \textbf{reduced bias}. Bias in AI systems can perpetuate harmful stereotypes or lead to inequitable outcomes, particularly in mental health care, where individualized and empathetic responses are critical. To reduce bias, the system must leverage a diverse range of data sources. By increasing the \textbf{number of nodes participating} in a federated learning framework, the chatbot can learn from a broader dataset, ensuring that its responses are not disproportionately influenced by any one population or context. This approach helps mitigate the risk of bias and enhances the system’s fairness and inclusivity.

Equally important, the chatbot must demonstrate \textbf{empathy} in its interactions. The ability to offer empathetic responses is central to mental health support, where users need to feel understood and validated. By using an \textbf{evidence-proved chatbot}, one that has been rigorously tested and validated against existing mental health interventions, the system can ensure that its responses are both scientifically grounded and empathetic. This not only enhances user satisfaction but also increases the chatbot’s effectiveness in delivering meaningful support.

Finally, \textbf{privacy} is a critical concern in any digital mental health intervention. Users must feel confident that their interactions with the chatbot remain private and secure. \textbf{Federated learning} offers a solution to this challenge by enabling the chatbot to learn from user interactions without storing sensitive data in a central location. This decentralized approach allows the system to improve over time while minimizing the risk of data breaches and ensuring that user privacy is respected at all stages.

In summary, the feasibility of an AI-enabled chatbot for mental health support depends on its ability to meet these five key criteria: trustworthiness, security, bias mitigation, empathy, and privacy. The integration of \textbf{human-in-the-loop oversight}, \textbf{cryptographic security}, \textbf{diverse data sources}, \textbf{evidence-based interactions}, and \textbf{federated learning} creates a robust framework that addresses the unique challenges of providing digital mental health support. The proposed system, as outlined in Figure \ref{FeaturseMap}, represents a comprehensive approach to designing a trustworthy, secure, reduced bias, empathetic, and privacy-preserving chatbot capable of addressing the mental health needs of diverse populations.

\section{Federated Learning as a Potential Solution }

\begin{figure}
  \centering
  \includegraphics[width=3in]{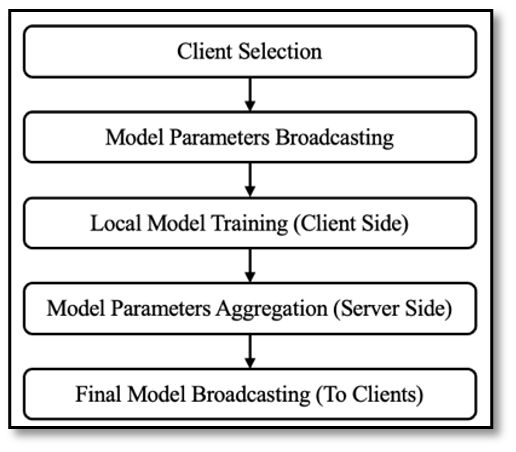}
  \caption{Federated Learning Steps}
  \Description{Federated Learning Steps}
  \label{FL}
\end{figure}

Google introduced Federated Learning in 2016, a recent trend in which a central server supervises a group of edge nodes (clients) that collaboratively learn a shared machine-learning (ML) model without sharing their sensitive data \cite{mammen_federated_2021, rieke_future_2020, suruliraj_federated_2022}. All decentralized participating nodes learn the shared ML model when a trusted curator aggregates their locally computed model’s parameters to a centralized server and then sends back the resulting model, as depicted in Figure ~\ref{FL}, preserving the privacy of their data \cite{geyer_differentially_2018, mcmahan_communication-efcient_2017, suruliraj_federated_2022}. Federated Learning has been a trend in different fields, especially healthcare, due to patient data-sensitive nature \cite{suruliraj_federated_2022}. Although studies have proposed frameworks asserting that federated learning is a promising approach to utilize and protect the privacy of patient data silos that reside on health institutional servers, it suffers from multiple challenges \cite{geyer_differentially_2018, mammen_federated_2021, suruliraj_federated_2022}.

Two broad classifications of federated learning challenges are training-related concerns and privacy and security concerns \cite{mammen_federated_2021}. Communication exhaustion or cost due to training iterations and the heterogeneity of data (non-identically distributed data) and devices used in the training process are instances of training-related concerns. Whereas membership inference attacks, data poisoning attacks, model poisoning attacks, byzantine attacks, and backdoor attacks are instances of privacy and security concerns \cite{li_federated_2020, mammen_federated_2021, wen_survey_2023, mcmahan_communication-efcient_2017}. Therefore, practical and effective techniques to alleviate the above problems are urgently needed when considering federated learning in building AI-enabled chatbots as mental health support.

For example, implementing a cryptographic technique in federated learning, such as homomorphic encryption, is required to preserve data privacy \cite{geyer_differentially_2018, munjal_systematic_2023}. Homomorphic encryption in federated learning can preserve data privacy by giving participating clients the ability to perform operations on ciphertext, producing outcomes like those gained when performing the same operations on the original plaintext \cite{munjal_systematic_2023, wen_survey_2023}. Also, Minimizing the communication rounds among edge nodes and the aggregator server as well as the size of messages sent in these rounds could solve the communication cost in federated learning \cite{li_federated_2020}. Another technique is applying \cite{de_choudhury_integrating_2018} suggestion of incorporating human intelligence with AI intelligence to validate model construction, spot unobserved data flaws, evaluate data bias, and proactively identify, control, and mitigate risk. The Human AI collaboration must include multi-stakeholder participation to assess all aspects of AI \cite{pandi-perumal_artificial_2024}.

In this paper, we propose a framework for building an AI-enabled chatbot that utilizes federated learning in training the AI model and magnifies the role of the human element in the loop. Stakeholders must assess the state of AI use for mental health. As \cite{graham_artificial_2019} advised, it is the responsibility of stakeholders, including all of us, to make an active move to transform mental health support by incorporating the mental health clinicians' expertise with the expertise of data scientists and other scientists. Our ultimate goal is to create a reliable, evidence-proved, reduced bias, and empathic AI-enabled chatbot that can serve as first aid for mental health, especially in underrepresented and underserved communities.
\begin{figure}[b]
  \centering
  \includegraphics[width=\linewidth]{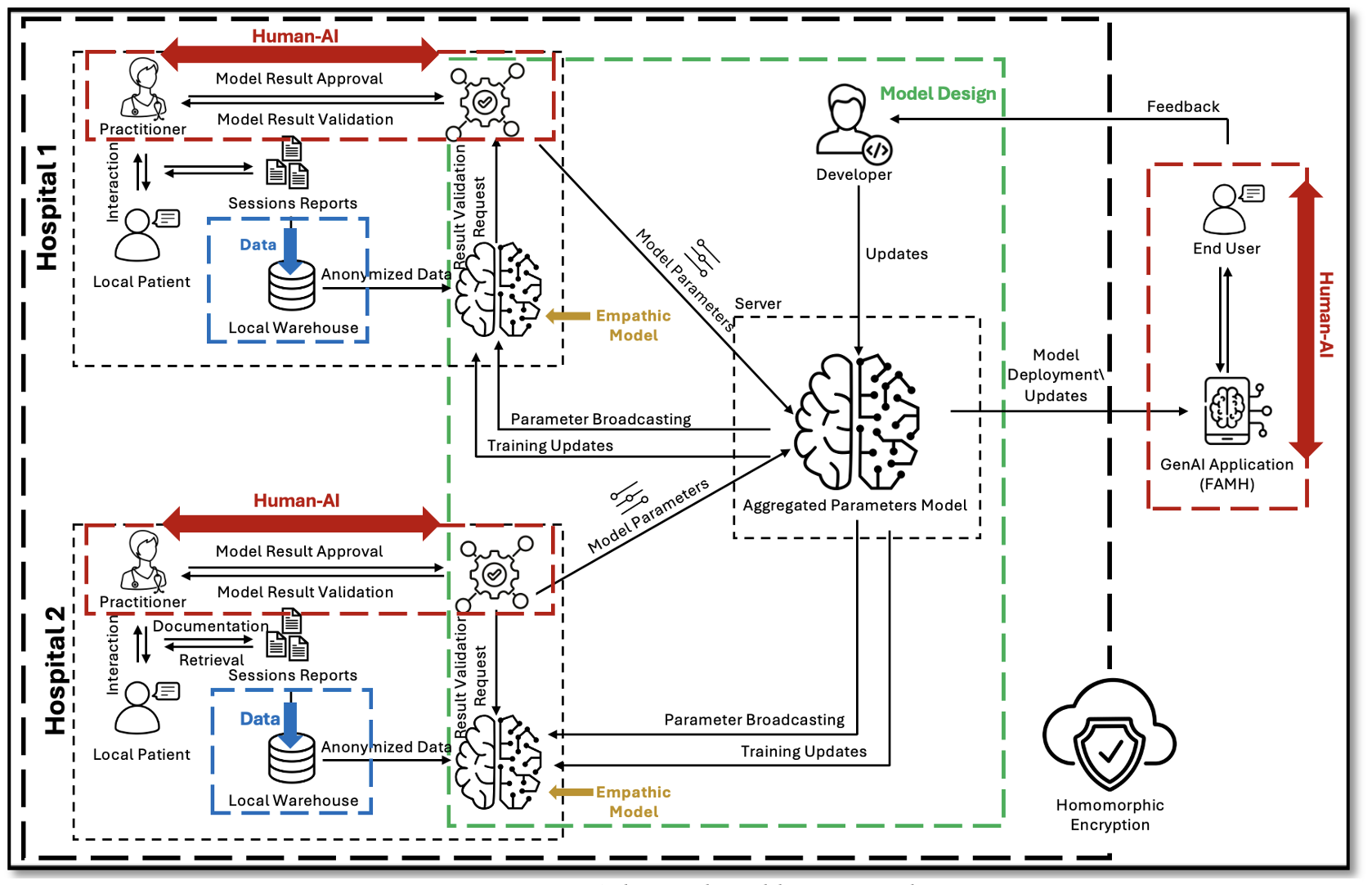}
  \caption{First Aid Mental Health Framework}
  \Description{First Aid Mental Health Framework}
  \label{FAMH}
\end{figure}
\section{First Aid Mental Health}

Although some researchers have raised concerns that AI-enabled chatbots may lack empathy and personalization, with these limitations highlighted as obstacles in previous studies \cite{balcombe_ai_2023, montemayor_principle_2022}, recent findings challenge this view. A study by \cite{welivita_are_2024} asserted that the empathic responses generated by large language models (LLMs) can, in fact, surpass those of humans. An empathic AI-enabled chatbot is grounded in empathy theory from psychology, where cognitive empathy, a learnable and teachable skill, plays a central role in counseling sessions between clinicians and patients \cite{hardee_overview_2003, lin_empathy-based_2023, platt_empathic_1994}. Building on this foundation, we aim to incorporate a model with proven superior empathy into our proposed framework. Specifically, \cite{welivita_are_2024} found that ChatGPT-4 outperformed its peers in a between-subject study evaluating the empathic quality of LLM-generated responses.

Figure ~\ref{FAMH} lays out the main stakeholders involved, the general techniques to be used, and the data and operation flow. We reassert \cite{de_choudhury_integrating_2018} point of view on the importance of human intelligence in the process, where constant, expert feedback is crucial for the success of the entire framework.

The proposed framework's success depends on the number of edge nodes (hospitals and clinics) participating in the federated learning process. Each participating node will enlarge the final dataset used and empower the machine learning model training process. Since federated learning depends on accumulating models’ parameters that were trained on unique datasets from multiple and distinct edge nodes, the resulting aggregated model on the centralized server will be trained on diversified data, ensuring more inclusive and unbiased digital mental health support. Having multiple parties (hospitals in our case) who collaboratively train local models with their local datasets and push the model's parameters to an aggregator in federated learning will minimize the bias problem and make the model more generalizable in mental health \cite{mammen_federated_2021, rieke_future_2020}.

Each participating hospital or clinic (for example, hospital 1 and hospital 2 in figure ~\ref{FAMH}) will create reports from doctor/patient sessions, anonymize them, and add them to a local database. It is important to emphasize that all datasets used should follow the same format to ensure the data heterogeneity problem is solved in federated learning. The local data engineer is responsible for confirming compliance with the agreed-upon data format. Moreover, it is important to highlight the role of mental health clinicians and data engineers in diversifying the input data to eliminate bias propagation problem \cite{chang_bias_2023} in federated learning. The local empathic machine-learning model will then be trained on these reports and wait for an aggregator request to push the model parameters only while preserving data privacy by not sharing them.

The model parameters will not be pushed to the aggregator until local data scientists and mental health clinicians from the participating edge node validate the local model's result. This step, which incorporates human intelligence, is crucial to making the resulting model more reliable. To eliminate false positive (FP) and false negative (FN) cases, clinician expertise must be seriously aligned with machine learning models. A local data scientist should test the chatbot's performance using evaluation metrics. For example, deepeval \cite{noauthor_introduction_2024} could be used to check the chatbot's answer relevancy, hallucination, and bias. This process must take place with the help of a local mental health clinician in order to formulate the different test cases. A local mental health clinician should also validate the model by interacting with it and providing different prompts and scenarios to examine the model responses, rationality, fluency, relevance, and coherence. A semantic textual similarity analysis, for example, using the Multilingual Text Enhancer (MLTE) approach \cite{feng_semantic_2024}, could also be performed to test the semantic similarity of the LLM response to the expected response. 

The pushed local model parameters will be used by an aggregating model on a centralized server. The model will enhance its performance and push back its parameters to all participating edge nodes to enhance their performance in turn. Data scientists (developers) will be involved in validating the performance of the central server's resulting model and providing constant updates. The whole process of federated learning will use a reliable and evidence-proven encryption technique. An example of encryption technique is the homomorphic encryption that allows performing operations without decrypting data in advance \cite{acar_survey_2019}. 

The resulting model from the federated learning approach will be used as a first-aid mental health support chatbot. The chatbot will always be available for those who need access to mental health services but suffer from mental healthcare barriers. No data will be collected from the users to enhance the chatbot's performance. To preserve users' privacy and the chatbot's reliability, the only data that will be used to enhance the chatbot's performance is from the participating hospitals. Developers could only collect survey feedback from users to improve the chatbot's features and tackle the pain points associated with using it.

\section{Conclusion}
The development of AI-enabled mental health support tools presents both opportunities and challenges. This paper has explored these aspects by analyzing the capabilities and limitations of large language models (LLMs) in replicating the therapeutic responses of human mental health professionals. Our analysis, which utilized cosine similarity metrics and qualitative assessment, revealed that although models like ChatGPT-4 and Mistral produce grammatically correct and coherent responses, they often lack the nuanced empathy and emotional depth characteristic of human therapists. This finding underscores a key limitation: while AI models can imitate surface-level responses, they struggle to fully engage with the emotional and psychological complexities of distressed individuals. Moreover, the cross-comparison between LLMs demonstrated that even state-of-the-art models exhibit variations in their output, further complicating their use in sensitive contexts such as mental health.

A key contribution of this paper is the novel framework we proposed for developing a more reliable, empathetic, and secure AI-enabled chatbot for first-aid mental health support. By integrating human oversight—via a “human-in-the-loop” approach—alongside a federated learning system, we aim to address crucial challenges such as bias mitigation, privacy, and trustworthiness. Federated learning ensures that the chatbot can learn from diverse datasets distributed across multiple healthcare nodes, without compromising user privacy. This decentralized approach not only helps reduce bias but also enhances the model’s generalizability across different population groups, making it a more inclusive solution for underrepresented and underserved communities.

One of the key findings of this study is that while LLMs have the potential to offer support, their effectiveness depends significantly on expert intervention. Our analysis highlights that incorporating human intelligence at various stages—model validation, data scrutiny, and feedback gathering—can help mitigate issues such as hallucinations, bias propagation, and shallow emotional engagement. The role of clinicians, in particular, is indispensable in ensuring that AI responses align with therapeutic best practices and are free from harmful stereotypes or inaccuracies.

Furthermore, this paper emphasizes the importance of using cutting-edge cryptographic techniques, like homomorphic encryption, to safeguard sensitive mental health data during federated learning. The integration of such privacy-preserving technologies, coupled with constant human oversight, creates a multi-layered framework that prioritizes both security and ethical considerations.

In conclusion, while LLMs offer a promising avenue for mental health support, it is evident that the current models are not yet ready to fully replace human therapists, especially in emotionally sensitive scenarios. However, with the right balance of advanced AI techniques, expert human involvement, and secure, privacy-preserving frameworks, AI-enabled chatbots could become a valuable tool in bridging the mental healthcare gap. The novel framework proposed in this paper presents a pathway toward developing an AI system that is reliable, reduced in bias, empathetic, and privacy-focused, capable of providing first-aid mental health support to diverse populations. As we move forward, a collaborative effort between mental health professionals, data scientists, and users is crucial to making this solution both effective and widely accepted.

\section{Future Work}
While large language models keep advancing and unlocking massive potentials, a possible research area would be using the proposed AI-enabled chatbot by mental health clinicians to assess their patients' conditions. If the proposed chatbot is used as a patient's mental health companion, it could analyze the patient's language patterns to detect early signs of treatment responses as well as make early psychological and pharmacological intervention recommendations. Besides, collecting feedback from the AI-enabled chatbot users through a controlled experiment would increase the feasibility of the model. This would help understand the most important features to incorporate for mental health support during a crisis and the sense of urgency. 


\bibliographystyle{ACM-Reference-Format}
\bibliography{references}

\begin{appendices}
\section{A Sample Prompt for A Patient Question, the Human Therapist's Response along with LLM's Outputs} \label{Appendix_Output}

\begin{alltt}
\textbf{Prompt:} You are a Registered Psychotherapist female. You received the following question from a person
suffering from a mental health issue. What would be your response? Do not provide an unnecessary
response. Reply as if you were talking to the person directly.  
The question is: "I just took a job that requires me to travel far away from home. My family and
I really need this job. People keep telling me I have "anxiety" and I'm terrified of having an
anxiety attack on the road. This is all new to me. What can I do?"

\textbf{Therapist Response:} Anxiety can cause such a feeling of discomfort in our bodies that it
can seem unbearable. The concern that I can hear in what you have written is that the thought of
having some anxiety has now left you uncomfortable with any anxiety at all - and that can become
a problem, indeed. As opposed to thinking about your anxiety as something much bigger than you,
it could be helpful to start breaking it down. What are the things specifically that cause you
anxiety about this new job? Is it being away from your family? If so, why?Once you start breaking
down what causes you to feel anxiety into smaller "mouthfuls", then we are far better able to
understand what has led to discomfort and start to problem solve the issue.

\textbf{ChatGPT Response:} It sounds like you’re going through a challenging time, and it's
understandable to feel worried about this new situation. Anxiety, especially when it’s unfamiliar,
can feel overwhelming, but there are ways to manage it.
First, recognize that anxiety is a normal response to new and uncertain situations. It can help to
prepare some simple strategies for when you feel it coming on. Breathing exercises, such as taking
slow, deep breaths, can help calm your nervous system. You might also try grounding techniques, like
focusing on your senses — noticing five things you can see, four you can touch, three you can hear,
two you can smell, and one you can taste.
Since you're worried about having an anxiety attack while traveling, plan some breaks or calming
activities. Listening to a favorite podcast, practicing mindfulness, or even calling someone you
trust could help you feel more grounded.
It could also be helpful to speak with a mental health professional about these feelings.
They can work with you to develop a personalized plan to manage anxiety effectively.
Remember, it’s okay to feel this way, and there are many ways to get support. You’re not alone.


\textbf{Mistral Response:} I'm really glad you've reached out to discuss this. It's completely normal to 
feel anxious when facing significant changes, like a new job with extensive travel. Let's focus on
some strategies to help you manage this anxiety.
Firstly, it's important to understand what an anxiety attack or panic attack feels like. Symptoms
can include a racing heart, shortness of breath, dizziness, and overwhelming fear. It's crucial
to remember that while these symptoms are unpleasant, they are not dangerous.
Here are some techniques you can use on the road:
1.	Deep Breathing: If you start to feel anxious, try taking slow, deep breaths. Inhale for a
count of four, hold for four, and exhale for four. This can help calm your body's physical
response to anxiety.
2.	Grounding Techniques: One example is the 5-4-3-2-1 method. Name 5 things you can see,
4 things you can touch, 3 things you can hear, 2 things you can smell, and 1 thing you can taste.
This helps bring your focus back to the present moment.
3.	Preparation: Plan your trips in advance. Knowing where you're going, how you're getting there,
and where you're staying can help reduce uncertainty and anxiety.
4.	Self-Care: Ensure you're getting enough sleep, eating healthily, and staying hydrated. Physical
health can greatly influence your mental state.
5.	Stay Connected: Regularly call or video chat with your family. Hearing their voices and seeing
their faces can be comforting.
6.	Professional Help: If your anxiety feels unmanageable, consider speaking to a mental health
professional. They can provide strategies and tools tailored to your needs. Some therapists even
specialize in remote or online therapy, which could be helpful given your travel schedule.
Lastly, remember that it's okay to have anxiety, and it's not a sign of weakness. It's a normal
human response to stress. With the right tools and support, you can manage it and succeed in
your new job. You're not alone in this. Let's take it one step at a time together.

\end{alltt}

\end{appendices}

\end{document}